\documentclass[a4paper,fleqn]{cas-dc}

\usepackage[numbers]{natbib}

\def\tsc#1{\csdef{#1}{\textsc{\lowercase{#1}}\xspace}}
\tsc{WGM}
\tsc{QE}
\tsc{EP}
\tsc{PMS}
\tsc{BEC}
\tsc{DE}
\usepackage{amssymb}
\usepackage{amsmath}
\usepackage{bm}
\usepackage{pdflscape}
\usepackage{caption}
\usepackage[utf8]{inputenc}
\usepackage[T1]{fontenc}
\usepackage{url}
\begin{document}
\let\WriteBookmarks\relax
\def\floatpagepagefraction{1}
\def\textpagefraction{.001}
\shorttitle{Bayesian triage of structural heart disease}
\shortauthors{M. J. Colebank}

\title[mode = title]{Uncertainty-aware classification and triage of structural heart disease using electrocardiography and echocardiography metrics}

\author[1]{Mitchel J. Colebank}[type=editor,
auid=000,bioid=1,
orcid=0000-0002-2294-9124]
\cormark[1]
\ead{mjcolebank@sc.edu}


\affiliation[1]{organization={Department of Mathematics, University of South Carolina},
	addressline={}, 
	city={Columbia},
	postcode={29208}, 
	state={South Carolina},
	country={USA}}

\cortext[cor1]{Corresponding author}

\begin{abstract}
Machine learning methods provide a methodological innovation that can help screen for cardiovascular disease through noninvasive and readily available measurement modalities. Recent investments in using electrocardiogram (ECG) data to screen for structural heart disease (SHD) are one example, where ECGs provide a low-cost, available modality for screening. This has led to the EchoNext dataset, a paired ECG-echocardiogram data repository for testing new methods of SHD detection. However, relatively few studies have investigated how more probabilistic classification through Bayesian inference may improve uncertainty quantification in this setting. Moreover, few studies have considered how triage systems can be developed to alleviate healthcare bottlenecks, such as the review of data from underserved, rural clinics by expert sonographers for SHD assessment. In this study, we leverage existing ECG-echocardiogram data to compare frequentist and Bayesian neural network classifiers. We show that the Bayesian approach is comparable or better than frequentist methods in SHD classification, and that they have a more robust uncertainty quantification attached to them. We provide an example of how this uncertainty-aware classification scheme can be used for screening SHD, providing a proof-of-concept for how machine learning can help with triage in getting individuals expert sonographer input when SHD is highly likely or measurements are highly uncertain.
\end{abstract}




\maketitle



\section{Introduction} \label{Intro}
Data science and machine learning methods have provided a new frontier in how healthcare data is leveraged for clinical decision making and patient monitoring \cite{Chua2022,Gandhi2018,Narula2016}. While data centered approaches have been around for quite some time, recent years have seen a drastic focus on how data-driven solutions may help with disease prognosis and patient-specific care. One target area is in the management of cardiovascular disease, the leading cause of death in the world \cite{AHA2026}, which can greatly benefit from the ability to integrate multi-modal data from various non-invasive and invasive tests. However, a majority of machine learning methods suffer from a critical limitation in their ability to describe and account for uncertainty, especially when measurements may be noisy or sparse \cite{Chua2022}.

This limitation is attributed to training of machine learning models through optimization, without explicit focus on how \textbf{statistical ideologies} are directly integrated into the machine learning architecture. The main reasoning for this is that many machine learning methods are over parameterized (e.g., in neural networks) such that any statistical interpretation of uncertainty may be difficult through classical (often called \textbf{frequentist}) frameworks \cite{Arbel2026,Ding2020}. A remedy for this is to transition from a frequentist framework, where parameters are assumed fixed but unknown, to a \textbf{Bayesian} framework \cite{Arbel2026,Elsayad2015,Ordovas2023}. In this latter perspective, parameters, $\bm{\theta}$ are treated as random variables with some unknown distribution for which we would like to infer. We construct our \textbf{posterior} beliefs about $\bm{\theta}$ through (i) our \textbf{prior} beliefs and (ii) the \textbf{likelihood} of the data, conditioned on our predictions from the model through the parameter values. In this setting, both the epistemic uncertainty (the reducible uncertainty due to our model) and the aleatoric uncertainty (the uncertainty due to limited or noisy observations) can be quantified. This provides a necessary improvement in how uncertainty is quantified and provides machine learning tools with the ability to express their inability to make predictions \cite{Arbel2026}. This is a necessary area of data science that needs to be addressed prior to implementations in the clinical setting.

In this manuscript, we focus on a specific application of these methods in the context of heart disease screening and risk assessment. Echocardiography is considered the gold standard for screening heart disease, including a broad category of ``structural heart disease'' (SHD), which can include left-sided, right-sided, or valvular heart diseases \cite{Poterucha2025}. Echocardiography is extremely useful in this context, but requires expert sonographer interpretation in order to assess metrics such as ejection fraction (EF, the percent of blood ejected out of the heart), as this is subject to measurement noise due to echo technological limitations and subject movement. This makes it especially difficult for clinics without expert sonographers (e.g., in the rural south United States), who often rely on sending results to expert cardiologists to review remotely \cite{Fazlalizadeh2024}. Alternative measurements, such as electrocardiography (ECG), are easier to implement across clinics, but provide less diagnostic power in comparison to echocardiography \cite{Elias2022,Poterucha2025,Ulloa-Cerna2022}. This provides an excellent testbed for machine learning, which could expedite the review process or provide a triage system for echocardiography assessment by expert sonographers \cite{Fazlalizadeh2024,Pedroso2026}. 

In this study, we compare standard neural network architectures in the frequentist and Bayesian setting in the context of SHD classification. We leverage an existing repository of ECG and echocardiography across 100,000 unique entries called the ``EchoNext'' dataset \cite{Elias2022,Poterucha2025}. We consider three subsets of data availability to identify how less uncertain, readily available measurements may be leveraged in detecting SHD. We compare these across frequentist and Bayesian settings, providing a comparison of how these models can be interpreted, and how uncertainty can be quantified in this setting.


\begin{table*}
	\centering
	\begin{tabular}{|c|c|}
		\hline
		Dataset & Features \\
		\hline
		Dataset 1 ($D_1$) & $D_1=\{$sex, age, PR, QRS, QT, Atr Con, Vent Con$\}$  \\
		Dataset 2 ($D_2$) & $D_2=\{D_1,$ Peri Eff, IVS, PWT$\}$\\
		Dataset 3 ($D_3$) & $D_3=\{D_2,$ TR Max Velocity, EF$\}$ \\
		\hline
	\end{tabular}
	\caption{Features included in each individual dataset, $D_1$, $D_2$, and $D_3$, in the present study. See Figure \ref{fig:data_distribution} for the data distribution. PR: PR interval from ECG; QRS: QRS  duration; QT: QT interval length; Atr Con: atrial contraction rate; Vent Con: ventricular contraction rate; Peri Eff: possible pericardial effusion; IVS: interventricular septal width; PWT: posterior wall thickness;  TR Max Velocity: maximum velocity of regurgitation at the tricuspid valve; EF: ejection fraction out of the left ventricle.}
	\label{tab:Dataset}
\end{table*}

\section{Data - EchoNext Dataset}
We focus on an existing dataset that includes typical non-invasive echocardiography measurements obtained during the screening for structural heart disease. The EchoNext Dataset \cite{Elias2022,hughes2026echonext,Poterucha2025} is an open-access dataset, available from Physionet \cite{PhysioNet}, that includes over 100,000 individual measurements from echocardiography imaging. Briefly, the data were retrospectively collected from adult patients 18 years or older who underwent electrocardiography (ECG) and echocardiography imaging. The latter were extracted from the Syngo Dynamics (Siemens) and Xcelera (Philips) systems. 

For the ECG, we include heart rate (HR, in beats per minute), PR interval (milliseconds), QRS  duration (milliseconds), and the QT interval length (milliseconds). The atrial contraction rate (Atr Con, in beats per minute) and ventricular contraction rate (Vent Con, in beats per minute) are also calculated and used. All ECG features were extracted from XML files obtained from GE MUSE Nx 10.2. The measurements from echocardiography included left ventricular ejection fraction (LVEF, the percent of blood ejected out of the ventricle), interventricular septal thickness (IVS, in centimeters), posterior wall thickness (PWT, in centimeters), pulmonary artery systolic pressure (PASP, mmHg) derived from Doppler measurements, and tricuspid regurgitation maximum velocity (TR Max Velocity, in meters per second). The maximum of IVS and PWT is considered to be the left ventricular wall thickness (LVWT) for disease classification purposes. Pericardial effusion (Peri Eff) which assesses fluid accumulation in the pericardial space is categorized as none, small, moderate, or large. We also include sex and age as additional demographic variables. 
To handle missing data for continuous variables, we used median imputation.

\begin{figure}
	\centering
	\includegraphics[width=1\linewidth]{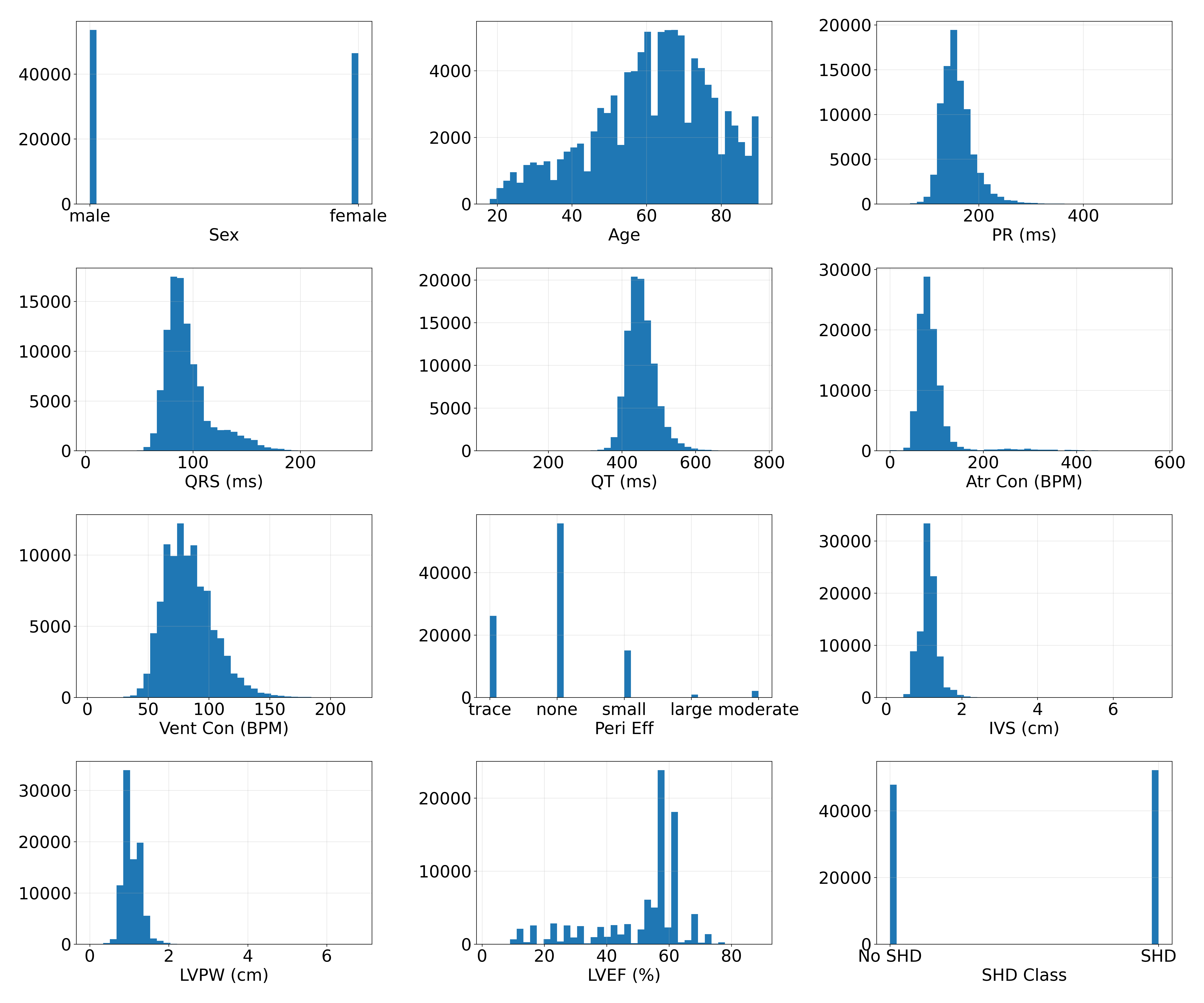}
	\caption{Distribution of values for covariates used in the three designs $D_1$, $D_2$, and $D_3$. PR: PR interval from ECG; QRS: QRS  duration; QT: QT interval length; Atr Con: atrial contraction rate; Vent Con: ventricular contraction rate; Peri Eff: possible pericardial effusion; IVS: interventricular septal width; PWT: posterior wall thickness;  TR Max Velocity: maximum velocity of regurgitation at the tricuspid valve; EF: ejection fraction out of the left ventricle; SHD: structural heart disease.}
	\label{fig:data_distribution}
\end{figure}

As described in the original EchoNext dataset \cite{Poterucha2025}, an individual is considered to have ``structural heart disease" if \textbf{any} of the following conditions are met within the moderate and severe disease categories:
\begin{enumerate}
    \item \textbf{Moderate Disease:} $\mathrm{LVWT} \geq 1.3\,\mathrm{cm}$, $\mathrm{LVEF} \leq 45\%$, $\mathrm{PASP} \geq 45\,\mathrm{mmHg}$, and $\mathrm{TR\ Max\ Velocity} \geq 3.2\,\mathrm{m/s}$.
    \item \textbf{Severe Disease:} $\mathrm{LVWT} \geq 1.6\,\mathrm{cm}$, $\mathrm{LVEF} \leq 35\%$, $\mathrm{PASP} \geq 60\,\mathrm{mmHg}$, and $\mathrm{TR\ Max\ Velocity} \geq 3.6\,\mathrm{m/s}$.
\end{enumerate}

We examined three distinct groupings of data to assess how our infrastructure classified datasets as indicative of structural heart disease. In the first dataset ($D_1$), we consider ECG metrics, sex and age. In the second dataset ($D_2$), we include the ECG and demographic data, but also add IVS and PWT measurements, as these are considered less subject to measurement error in comparison to velocity and pressure-related metrics \cite{Gandhi2018}. We emphasize that this dataset is a ``limited-echo'' dataset, which includes some descriptors that should increase the ability of the machine learning models to predict SHD. In our final dataset ($D_3$), we include all ECG and a larger subset of echocardiography, including EF and TR Max Velocity. This final dataset is provided as a methodological test only, since the above metrics from the echocardiography provide nearly all the data needed to predict whether or not SHD is present. Thus, $D_3$ is effectively a guaranteed high-predictive dataset that should only be interpreted as an upper bound on predictive capabilities. Table 1 provides a summary of this information. Histograms describing the data distributions can be found in Figure \ref{fig:data_distribution}.

\section{Methods}
\subsection{Neural Networks}
We consider the class of function approximators broadly defined as neural networks. Given their extensive use in multiple domains for handling large datasets \cite{Chua2022,Elias2022}, we consider them for classifying data indicative of structural heart disease using the three datasets described above. We provide a mathematical definition of the neural network to help clarify how frequentist and Bayesian inference settings differ. 

Consider some input data $\bm{X}\in\mathbb{R}^{N\times D}$ (real numbers) with $N$ samples of the $D$ dimensional observation vector $\bm{x_i}\in \bm{X}$. Suppose we are interested identifying the discrete class of the data source $y_i\in \Gamma = \{1,2,\dots, C\}, \ \  i=1,2,\dots,N$ with $C\in \mathbb{N}$ (integers) possible classes for each observation. The one-hot encoding of this data, $\bm{Y}\in\mathbb{R}^{C\times N}$ represents these classes concisely in a single matrix. Then an $L$ layer feed-forward neural network approximates the mapping from inputs $X$ to output class $Y$ through the recursive relationship
\begin{align}\label{eq:NN}
    &f(\bm{x}_j) = \phi_L\!\left(\bm{W}_L \,Z_{L-1}\!+ \bm{b}_L\right) \ \ \ \\
	&Z_{i} = \phi_{i}\!\left(\bm{W}_{i}Z_{i-1} + \bm{b}_{i}\right), \ \ i=2,3,\dots,L-1 \nonumber \\ 
	&Z_1 = \phi_1\!\left(\bm{W}_1 \bm{x}_i + \bm{b}_1\right) .\nonumber
\end{align}
In the above formulation, each layer of the neural network includes a weight matrix $\bm{W}_i$ and bias vector $\bm{b}_i$, which form the linear combination of the inputs, while $\phi_i$ is the nonlinear activation function. In the case of frequentist inferred parameters using gradient descent, we use ReLU activation functions for internal layers with a sigmoidal activation layer at the output for classification. We compared $\tanh$ to ReLU (see Supplementary material) for the frequestist setting and achieved nearly identical results. The cross-entropy loss is minimized in this setting. When constructing the area under the receiver operator curve (AUROC) and area under the precision-recall curve (AUPRC), we use bootstrap resampling of the test data after training the neural network to provide confidence intervals in the predictions \cite{Tyralis2024}.

\subsection{Bayesian Neural Networks}
In contrast to the frequentist approach where we seek optimal values of the weight and bias parameters, the Bayesian framework instead seeks probabilistic properties of the parameters \cite{Arbel2026,Jospin2022}. Let $\bm{\theta}$ denote the set of weights and biases in the neural network. The \textit{prior}, $p(\bm{\theta})$, describes our beliefs of the parameters before observing any data, while the \textit{likelihood}, $p(\bm{Y}|\bm{\theta})$, describes how likely the observations area conditioned on the current value of the weights and biases. We can relate these to our posterior beliefs in the parameters through Bayes' theorem
\begin{equation}
    p(\bm{\theta}|\bm{Y}) = \frac{p(\bm{Y}|\bm{\theta})p(\theta)}{\displaystyle\int_\Theta p(\bm{Y}|\bm{\theta})p(\theta)d\theta}.
\end{equation}
In the context of neural networks, the denominator (called the \textit{evidence}) is extremely non-convex and complicated in structure, thus limiting the ability to approximate the evidence without numerous samples. We discuss ways to make this integration tractable in the next subsection.

Once the posterior distribution has been effectively sampled from, we can construct the marginal, which represents our distribution in \textbf{output space} given the posterior density for the input parameters. The marginal is defined as
\begin{equation}\label{eq:marginal}
    p(\bm{y^*}|\bm{x^*,y,\theta}) = \int_\Theta p(\bm{y^*}|\bm{x}^*,\bm{\theta}')p(\bm{\theta}'|\bm{{y}})d\bm{\theta}'
\end{equation}
which effectively describes the uncertainty around some data point $\{\bm{y^*,x^*}\}$ conditioned on our training data  $\{\bm{y,x}\}$ and posterior beliefs in the parameters. 

The choice of prior distribution helps regularize the problem if chosen to be informative, e.g. using a Gaussian prior, $p(\theta)\sim\mathcal{N}(0,\Sigma)$, where $\Sigma$ is the prior covariance matrix \cite{Arbel2026,Jospin2022}. The zero mean Gaussian prior is effectively the same as using $\ell_2$ regularization in standard frequentist inference for neural networks \cite{Arbel2026,Jospin2022}. In the case of independent priors with the same prior variance $\sigma^2$ we get a diagonal covariance of the form $\Sigma = \sigma^2\bm{I}$, where $\bm{I}$ is the identity matrix. This then penalizes parameter values proposed far from the prior in favor of values in favor of the prior location unless the likelihood function strongly suggests otherwise. Throughout we use $\tanh$ activation functions, as these significantly outperformed ReLU (see Supplementary Material).

\subsection{Variational Inference}
Determining the posterior distribution can be computationally intensive because the evidence in the denominator is difficult to determine. For relatively shallow neural networks and small to moderate datasets (e.g., 500-5,000 data instances), one can employ Markov chain Monte Carlo (MCMC) techniques, such as Hamiltonian Monte Carlo (HMC). In the case of larger datasets, techniques like HMC become computationally intensive, requiring instead approximations of the posterior through reasonable approximation. This latter idea gives rise to variational inference. Given our interest in testing the model on large (70,000) samples, we use variational inference throughout.


Variational inference introduces a variational distribution, $q_\phi(\theta)$, which \textit{approximates} the posterior through its variational parameters $\phi$. The goal is to minimize the Kullback-Leibler (KL) divergence between the two
\begin{align}\nonumber
    &D_{KL}\left(q_\phi(\bm{\theta})||p(\bm{\theta}|\bm{Y})\right) = \int_\Theta q_\phi(\bm{\theta}')\log\left(\frac{q_\phi(\bm{\theta}')}{p(\bm{\theta}'|\bm{Y})}\right)d\bm{\theta}' \nonumber \\
    &= \mathbb{E}_q\left[\log\left(\frac{q_\phi(\bm{\theta})}{p(\bm{\theta}|\bm{Y})}\right)\right] \nonumber\\
    &=\log(p(\bm{Y})) - \left(   \mathbb{E}_q\left[\log \left(p(\bm{Y},\bm{\theta}\right)\right]    -
    \mathbb{E}_q\left[\log \left(q_\phi(\bm{\theta})\right)\right]\right). \nonumber
\end{align}
where we've used the definition from Bayes' theorem as well as the fact that the KL divergence is nonnegative. The notation $\mathbb{E}_q$ denotes the expectation with respect to the variation distribution $q_\phi(\bm{\theta})$. Since the KL divergence is strictly positive, we can establish the \textit{evidence lower bound} or ELBO
\begin{align}
    \textrm{ELBO}(q_\phi) &=  \mathbb{E}_q\left[\log \left(p(\bm{Y},\bm{\theta}\right)\right]    - 
    \mathbb{E}_q\left[\log \left(q_\phi(\bm{\theta})\right)\right] \nonumber \\
    &= \mathbb{E}_q\left[\log \left(p(\bm{Y}|\bm{\theta}\right)p(\bm{\theta}))\right]    - 
    \mathbb{E}_q\left[\log \left(q_\phi(\bm{\theta})\right)\right].\nonumber 
\end{align}

Given that $D_{KL}(q_\phi(\bm{\theta})||p(\bm{\theta}|\bm{Y})=\log(p(\bm{Y}))-\mathrm{ELBO}(q_\phi)$, our goal is to \textit{maximize} the $\mathrm{ELBO}$ by finding sufficient parameters of the variational distribution $q_\phi(\theta)$. Given that this is now an optimization problem over the variational distribution parameters $\phi$.

In contrast to frequentist training of neural networks, the Bayesian predictive distribution (the marginal defined in eq. \eqref{eq:marginal}) provides a more direct assessment of uncertainty in predicting the classes. In theory, once can use the full output distribution to make classification decisions (as discussed later for our ``triage'' scenario). Here, we use the average of the marginal for decision making, with a probability threshold of 0.5 used unless specified.

\subsection{Assessment Metrics}
We compare the two paradigms across different experimental designs and neural network architectures using standard metrics for classification. This includes the Receiver Operating Curve (ROC) and corresponding area under the curve (AUC). We provide 95\% credible intervals for these values in the Bayesian setting. The precision-recall curve and corresponding area underneath this curve (AUPRC) are also used to ensure any class-imbalances are mitigated. Confusion matrices are also provided for the classification error. 

To contrast uncertainty metrics, we compute the Brier Score \cite{Arbel2026,Chua2022,Jospin2022} (effectively the mean square error  for our models, given by
\begin{equation}
    \textrm{BS} = \frac{1}{N}\sum_{i=1}^N (\hat{p}_i-y_i)^2
\end{equation}
where $N$ is the number of data in the test set and $\hat{p}_i$ is the predicted probability for SHD from the neural network for input $x_i$. We also use the expected calibration error (ECE), defined as the error between the observations and model for discretized bins $b$ with size $n_b$ of predicted probabilities. This metric is calculated as
\begin{equation}
    \textrm{ECE} =  \sum_{b=1}^{B}\frac{n_b}{N}\mid\textrm{acc}(b) - \textrm{conf}(b)\mid
\end{equation}
where $N$ is the total number of data points, $\textrm{acc}(b)=(TP(b)+TN(b))/N(b)$ is the accuracy of the model using samples \textit{within the bin $b$}, with $TP$ being the true positives and $TN$ being the true negatives, and $\textrm{conf}$ is the confidence of the model (representing the probabilistic predictions) for the same bin. An ECE value of zero means that the true and predicted probabilities are inline, while large ECE values suggest over- or under-confident predictions (lower is better). We evaluate the ECE across 10 uniformly spaced bins.

We report risk-coverage trade offs for our model architectures \cite{Kompa2021,Tyralis2024}, as this can help better screen models when comparisons between standard metrics (e.g., AUROC) do not incorporate uncertainty in decision making. The coverage is defined by increasing percentages of data as ranked by their uncertainty. For instance, 10\% coverage would represent the 10\% of data that is most certain, while 90\% coverage includes all but the most uncertain 10\% of data. The risk is calculated as the average of the classification error for a given coverage set of data $c_k$ with coverage percentage $k$
\begin{equation}
    \textrm{Risk}(c_k) = \frac{1}{\textrm{size}(c_k) }\sum_{i\in c_k}|y_i-\hat{y}_i|
\end{equation}
where the data $y_i$ and predicted class $\hat{y}_i$ are computed for the data within the coverage set. We expect low risk for low coverage, but can then examine how much risk a model accrues as we incorporate more and more of the data that may induce larger uncertainty. In general, as we include more of the uncertain data, our Risk should increase.

The risk-coverage curve reflects the nature of the selective prediction very well by definition as the motivation of the selective prediction is to reduce the coverage of the model in order to achieve higher accuracy.

\subsection{Triage: Providing bounds for ``inconclusive'' decision making}
A potential benefit of machine learning tools is the ability to reduce waiting times for expert sonographers when a patient's medical data clearly suggests either a healthy heart (no SHD) or the presence of SHD. However, this tool must be able to acknowledge borderline cases, where there may be a chance of SHD that cannot be deduced by the selective measurements available. We thus consider a selective prediction protocol, where we build a decision making platform that can be ``inconclusive'' in the case of large uncertainty. To do this, the training of the neural network is done as normal in the Bayesian setting. Rather than use the mean of the marginal to predict whether we classify the data as ``No SHD'' or ``SHD'', we impose an uncertainty-aware system that can help us triage datasets.

In the case of a triage system, we want to enforce a fairly conservative scheme that can acknowledge ambiguity when the two class probabilities are not clearly distinguished. While this can be optimized based on clinical understanding, we take up a relatively simple threshold for inconclusive predictions. The anticipated predicted class is based on a threshold of 0.5, and we use compare average of the predictive distribution to this threshold. We then compare the credible interval of the SHD class to two thresholds. If the posterior mean is less than 0.5, but the 95\% credible interval includes 0.2, we run a risk of false-negative and missing the SHD for expert sonographer input. This is an extremely conservative framework that would rather be inconclusive than miss a possible SHD subject. Thus, we only predict the ``No SHD'' class if the 95\% credible interval for ``SHD'' is strictly smaller than the 0.2 threshold. Alternatively, we only want to confirm SHD in the absence of the expert sonographer if we are very confident in the prediction of SHD. This is enforced by only predicting SHD with confidence if the 95\% does not include 0.5 (i.e., the posterior is localized away from the decision boundary). Any predictions from the model that do not make a confident prediction (SHD upper bound includes 0.2 or SHD lower bounds includes 0.5) are deemed ``inconclusive'', and would require expert sonographer input. This can be written concisely as
\begin{equation}
    \hat{y} = \begin{cases}
        \textrm{No SHD}, & \mathbb{E}\left[p_{\mathrm{SHD}}(x)\right]<0.5  \ \& \  Q_{0.975}\left(p_{\mathrm{SHD}}(x)\right)<0.2 \\
        \textrm{SHD}, & \mathbb{E}\left[p_{\mathrm{SHD}}(x)\right]\geq0.5  \ \& \  Q_{0.025}\left(p_{\mathrm{SHD}}(x)\right)>0.5 \\
        \textrm{Inconc.},  \ &\textrm{else}
    \end{cases} \nonumber
\end{equation}
where $Q_{0.975}$ and $Q_{0.025}$ denote the upper and lower quantile for the credible interval, respectively, and $p_{\mathrm{SHD}}(x) = p(\textrm{SHD}|x)$. The above considers data as inconclusive (``Inconc'') when uncertainty bounds satisfy our criteria.

\section{Comparison study}
Our goal in this study is to showcase how neural network classifiers can be leveraged in both large and small data instances through the use of Bayesian inference. Using the three experimental designs in Table 1 ($D_1,D_2,$ and $D_3$), we compared frequentist and Bayesian neural network training across various network depth, width, and training data size. Our hypothesis is that \textbf{Bayesian frameworks handle uncertainty in a more robust manner than frequentist approaches, even when using dropout during training}. Table \ref{tab:Design} reflects the designs for our comparison. 

We compare our results for smaller datasets (a subsample of the original training and testing data) for two reasons. First, given that SHD is a broader category of disease, other specific forms of disease (e.g., congenital heart defects) may have a substantially smaller dataset that includes both echocardiography and ECG data. Second, clinical partners seeking to use machine learning methods may be restricted in how patient data may be used outside the clinic. This requires that we assess whether these methods can be implemented on smaller datasets, reflecting a population that might be obtainable in a much smaller hospital setting.


\begin{table*}
\centering
    \begin{tabular}{|c|c|c|c|c|}
    \hline
     \multicolumn{5}{|c|}{Frequentist Neural Network} \\
    \hline \hline
    Act. Func. & \# of neurons & \# of layers & dropout rate & \# Train/Test \\
    \hline
    ReLU & 10 & 3/5/10 & 0.20/0.50 & 8000/2000 \\
    ReLU & 20 & 3/5/10 & 0.20/0.50 & 8000/2000 \\
    ReLU & 10 & 3/5/10 & 0.20/0.50 & 70,000/20,000 \\
    ReLU & 20 & 3/5/10 & 0.20/0.50 & 70,000/20,000 \\
    \hline \hline
   
        \multicolumn{5}{|c|}{Bayesian Neural Network} \\
    \hline \hline
    Act. Func. & \# of neurons & \# of layers & prior variances & \# Train/Test \\
    \hline
    $\tanh$ & 10 & 3/5/10 & 0.25/0.1 & 8000/2000 \\
    $\tanh$ & 20 & 3/5/10 & 0.25/0.1 & 8000/2000 \\
    $\tanh$ & 10 & 3/5/10 & 0.25/0.1 & 70,000/20,000 \\
    $\tanh$ & 20 & 3/5/10 & 0.25/0.1 & 70,000/20,000 \\
    \hline
    \end{tabular}
    \caption{Design of the neural network architectures in this study.}
    \label{tab:Design}
\end{table*}

\section{Results}
The frequentist neural network training was conducted using a batch-size of 32 over 100 epochs using the Adam optimizer with a learning rate of $10^{-3}$. The frequentist network uses ReLU activation functions between layers and a sigmoidal function at the output layer for classification, with dropout layers in between. We leverage the PyTorch package in Python for this setting \cite{paszke2019pytorch}. For the Bayesian framework, we draw 2000 posterior samples with 10,000 steps during stochastic variational inference and using the Adam optimized with a learning rate of $10^{-3}$. We use a hyperbolic tangent between layers, and impose zero-mean, Gaussian priors for all the weights and biases with a common, prescribed variance.  This is done within the NumPyro package in Python \cite{bingham2019pyro,phan2019composable}. We used an AutoNormal guide in Numpyro, corresponding to a mean-field Gaussian variational approximation over the latent neural-network parameters with the classification labels modeled using a categorical likelihood via the neural network output logits.

\subsection{Results for large dataset of 70,000}

Results from the frequentist-based neural network are provided in Table \ref{tab:fnn_summary_70k}, including the area under the precision-recall curve (AUPRC), Brier score, and ECE for each variation in network depth, network width, and dropout rate. As expected, the biggest difference exists between the experimental designs $D_1,D_2,$ and $D_3$. Using $D_3$ provides both TR Vel and EF (highly informative metrics), which subsequently allows the model to achieve the highers AUPRC values and lower Brier scores overall. Interestingly, the ECE scores are larger for $D_3$ relative to $D_2$ and $D_1$ overall. We observe mixed results in regard to the dropout rate, with the 5-layer model with 20 neurons having a notable improvement with greater dropout rates, while the 3- and 10-layer models tend to do as well or worse with more dropout. In general, AUPRC values are better for 20 neurons per layer, while Brier and ECE scores are comparable or better when 10 neurons are used. 

We provide similar results for the Bayesian neural networks in Table \ref{tab:bnn_summary_70k}. As was the case for the frequentist framework, classification using $D_3$ provides a superior AUPRC and Brier score compared to $D_2$ and $D_1$. Here, we see that the ECE scores are more comparable across designs in contrast to the frequentist results. In general, the AUPRC scores are better for the Bayesian models for comparable depth and width networks relative to the frequentist results. The effect of the prior distribution is minimal on the AUPRC, Brier, and ECE scores. The effect of network width (10 versus 20 neurons) is also minimal.

\begin{figure*}
    \centering
    \includegraphics[width=0.8\linewidth]{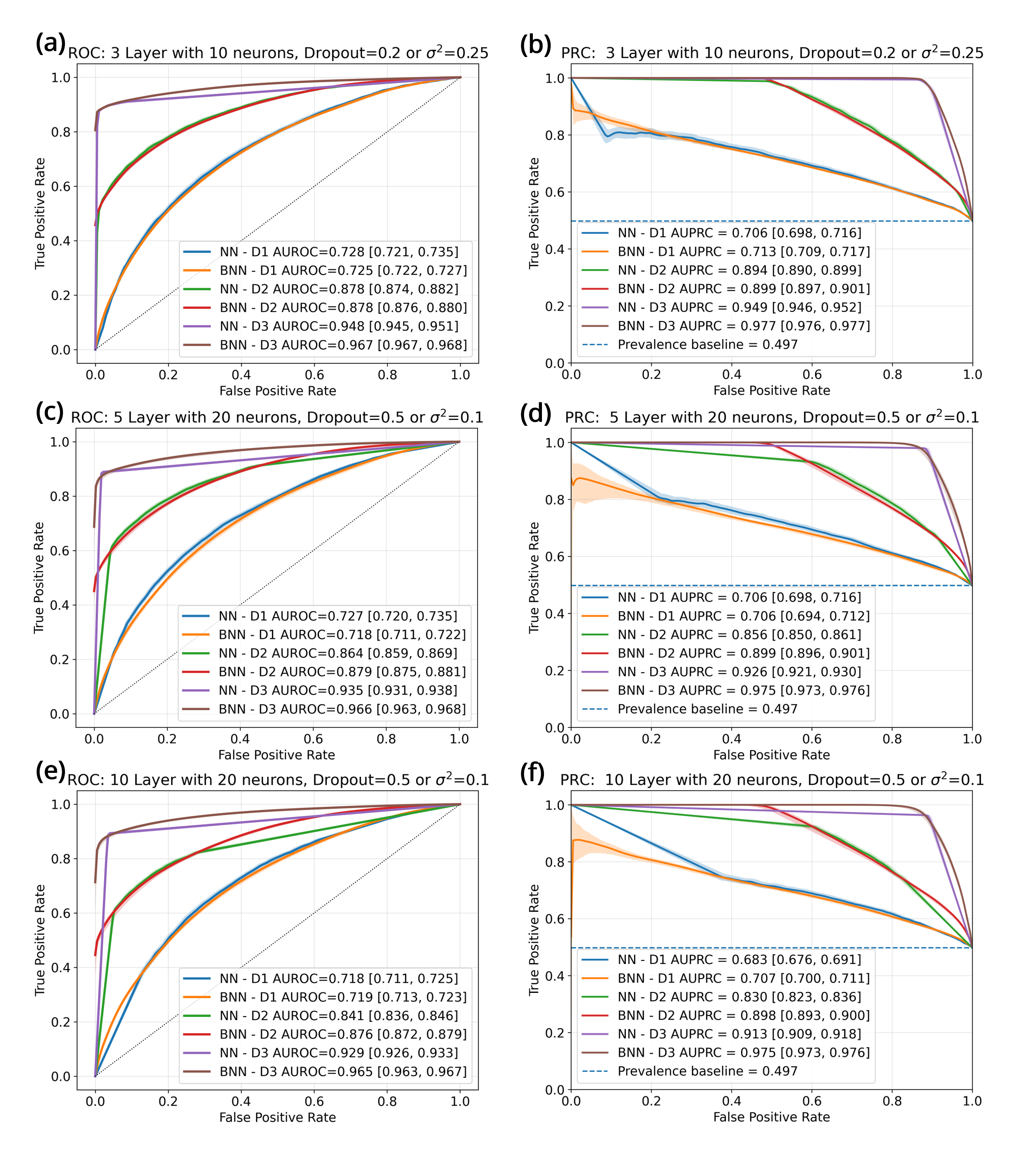}
    \caption{(Left) Receiver Operating Characteristic (ROC) curve for neural networks under the frequentist and Bayesian paradigm. Frequentist ROCs include 95\% confidence intervals while Bayesian ROCs include 95\% credible intervals, both also including the mean. Training data include 70,000 measurements, and results are shown for the 20,000 testing data points. (Right) Precision-Recall Curves (PRCs) for the corresponding ROC results.}
    \label{fig:ROC_PRC_70k}
\end{figure*}

We compared the two frameworks in terms of AUROC and AUPRC, as shown in Figure \ref{fig:ROC_PRC_70k}(a), (c), and (e), by selecting comparable frequentist and Bayesian trained networks based on network depth, dropout and prior variance, and width. Both frequentist and Bayesian results are shown with 95\% confidence and credible intervals, respectively, for the ROC curves by drawing 500 bootstrap samples or realizations from the predictive posterior. In general, we see improved performance using the Bayesian approach across neural network depth, width, and available data in contrast to the frequentist approach. The exception is for smaller network architectures using $D_1$, for which the frequentist approaches tend to do better. As previously stated, predictions with $D_3$ provide an upper bound for possible performance, and $D_2$ is clearly less informative but substantially stronger than the data provided in $D_1$. We see that the 3-layer, 10 neuron models tend to out perform the 5- and 10-layer models for both frequentist and Bayesian methods. The PRC and AUPRC results are shown in Figure \ref{fig:ROC_PRC_70k}(b), (d), and (f), and again show a clear improvement for using Bayesian versus frequentist methods. This is unsurprising, given that both classes are nearly equally represented. Conclusions from the ROC results are consistent with PRC curves. We also observe slightly wider confidence intervals for frequentist ROCs and PRCs than the Bayesian approach.

Given that uncertainty is a focus of our study, we provide risk assessments on the neural network models as well. Table \ref{tab:combined_20n_risk_only} shows the results across the different models for a fixed 20 neuron width with variable dropout, prior distribution variance, depth, and design. For $D_3$, we see small risk regardless of coverage level, with the Bayesian approach outperforming the frequentist approach across designs. When using data from $D_2$,  we still see lower risk for the Bayesian setting compared to the frequentist, but see a significant jump in risk (< 4\% to $\approx$10-14\%) at the 67\% coverage level. For the least informative design $D_1$, frequentist networks has similar or lower risk for increasing coverage compared to Bayesian methods. Risk metrics do not vary significantly with network structure, and are largely changing with coverage level and experimental design.


\begin{table*}[htbp]
\centering
\caption{Summary performance metrics for frequentist neural networks trained with 70,000 samples and tested on 20,000 samples. Columns are grouped by neuron count and specified as (dropout, number of layers ($N_L$)). Best values for each metric across all configurations are shown in bold. Higher AUPRC is better, while lower Brier score and ECE are better.}
\label{tab:fnn_summary_70k}
\begin{tabular}{lcccccc}
\hline
Metric | (Dropout, $N_L$)
& (0.20, 3) & (0.20, 5) & (0.20, 10)
& (0.50, 3) & (0.50, 5) & (0.50, 10) \\
\hline
\multicolumn{7}{c}{\textbf{20 neurons}} \\
\hline
AUPRC D3
& 0.9515 & 0.9407 & 0.9352
& 0.9531 & 0.9259 & 0.9133 \\
AUPRC D2
& 0.8937 & 0.8734 & 0.8792
& 0.8905 & 0.8557 & 0.8296 \\
AUPRC D1
& 0.7077 & 0.7047 & 0.7033
& 0.7092 & 0.7064 & 0.6829 \\
\hline
Brier D3
& 0.1029 & 0.1032 & 0.1025
& 0.1033 & 0.1032 & 0.1057 \\
Brier D2
& 0.1611 & 0.1607 & 0.1601
& 0.1605 & 0.1607 & 0.1632 \\
Brier D1
& 0.2128 & 0.2137 & 0.2133
& 0.2123 & 0.2130 & 0.2167 \\
\hline
ECE D3
& 0.2060 & 0.2047 & 0.2056
& 0.2053 & 0.2031 & 0.1979 \\
ECE D2
& 0.1095 & 0.1072 & 0.1151
& 0.1137 & 0.1110 & 0.0992 \\
ECE D1
& 0.0287 & 0.0299 & 0.0322
& 0.0234 & 0.0222 & 0.0197 \\
\hline
\multicolumn{7}{c}{\textbf{10 neurons}} \\
\hline
AUPRC D3
& 0.9487 & 0.9360 & 0.9291
& 0.9507 & 0.9256 & 0.8929 \\
AUPRC D2
& 0.8943 & 0.8855 & 0.8667
& 0.8985 & 0.8260 & 0.8232 \\
AUPRC D1
& 0.7063 & 0.7064 & 0.6990
& 0.7048 & 0.7008 & 0.6702 \\
\hline
Brier D3
& 0.1026 & 0.1027 & 0.1030
& 0.1034 & 0.1031 & 0.1135 \\
Brier D2
& 0.1608 & 0.1613 & 0.1620
& 0.1600 & 0.1643 & 0.1648 \\
Brier D1
& 0.2128 & 0.2132 & 0.2145
& 0.2130 & 0.2155 & 0.2195 \\
\hline
ECE D3
& 0.2072 & 0.2044 & 0.2035
& 0.2045 & 0.2023 & 0.1866 \\
ECE D2
& 0.1119 & 0.1132 & 0.1144
& 0.1126 & 0.1020 & 0.0987 \\
ECE D1
& 0.0272 & 0.0282 & 0.0256
& 0.0304 & 0.0245 & 0.0348 \\
\hline
\end{tabular}
\end{table*}

\begin{table*}[htbp]
\centering
\caption{Summary performance metrics for Bayesian neural networks trained with 70,000 samples and tested on 20,000 samples. Columns are grouped by neuron count and specified as (prior variance, number of layers ($N_L$)). Best values for each metric across all configurations are shown in bold. Higher AUPRC is better, while lower Brier score and ECE are better.}
\label{tab:bnn_summary_70k}
\begin{tabular}{lcccccc}
\hline
Metric | ($\sigma^2$, $N_L$) 
& (0.25, 3) & (0.25, 5) & (0.25, 10)
& (0.10, 3) & (0.10, 5) & (0.10, 10) \\
\hline
\multicolumn{7}{c}{\textbf{20 neurons}} \\
\hline
AUPRC D3
& 0.9762 & 0.9765 & 0.9754
& 0.9771 & 0.9766 & 0.9759 \\
AUPRC D2
& 0.9018 & 0.9030 & 0.9017
& 0.9010 & 0.9017 & 0.9009 \\
AUPRC D1
& 0.7219 & 0.7166 & 0.7088
& 0.7150 & 0.7121 & 0.7100 \\
\hline
Brier D3
& 0.0546 & 0.0544 & 0.0569
& 0.0555 & 0.0560 & 0.0570 \\
Brier D2
& 0.1364 & 0.1359 & 0.1377
& 0.1370 & 0.1373 & 0.1375 \\
Brier D1
& 0.2110 & 0.2123 & 0.2136
& 0.2121 & 0.2131 & 0.2145 \\
\hline
ECE D3
& 0.0103 & 0.0159 & 0.0116
& 0.0192 & 0.0226 & 0.0224 \\
ECE D2
& 0.0158 & 0.0266 & 0.0274
& 0.0168 & 0.0319 & 0.0201 \\
ECE D1
& 0.0246 & 0.0157 & 0.0208
& 0.0219 & 0.0151 & 0.0279 \\
\hline
\multicolumn{7}{c}{\textbf{10 neurons}} \\
\hline
AUPRC D3
& 0.9768 & 0.9761 & 0.9771
& 0.9762 & 0.9761 & 0.9765 \\
AUPRC D2
& 0.9004 & 0.9006 & 0.8991
& 0.9007 & 0.9006 & 0.9008 \\
AUPRC D1
& 0.7163 & 0.7197 & 0.7114
& 0.7166 & 0.7110 & 0.7115 \\
\hline
Brier D3
& 0.0531 & 0.0547 & 0.0540
& 0.0567 & 0.0560 & 0.0552 \\
Brier D2
& 0.1376 & 0.1371 & 0.1385
& 0.1373 & 0.1373 & 0.1373 \\
Brier D1
& 0.2113 & 0.2118 & 0.2133
& 0.2119 & 0.2134 & 0.2143 \\
\hline
ECE D3
& 0.0164 & 0.0105 & 0.0207
& 0.0250 & 0.0109 & 0.0290 \\
ECE D2
& 0.0167 & 0.0198 & 0.0127
& 0.0186 & 0.0181 & 0.0179 \\
ECE D1
& 0.0156 & 0.0277 & 0.0297
& 0.0219 & 0.0285 & 0.0329 \\
\hline
\end{tabular}
\end{table*}


\begin{table*}[htbp]
\centering
\caption{Selective-risk comparison for 20-neuron Bayesian (BNN) and frequentist (FNN) neural networks trained with 70K samples. Entries are reported in the order $D_3/ D_2/ D_1$. Lower risk indicates fewer errors among the retained least-uncertain predictions at the stated coverage. Bayesian models are indexed by prior variance $\sigma^2$, and frequentist models by dropout rate. $N_L$ is the number of layers.}
\label{tab:combined_20n_risk_only}
\small
\begin{tabular}{llcccc}
\hline
Framework & Tuning & $N_L$ & Risk @ 6\% & Risk @ 27\% & Risk @ 67\% \\
\hline
BNN & $\sigma^2=0.25$ & 3  & 0.000 / 0.000 / 0.140 & 0.000 / 0.005 / 0.234 & 0.019 / 0.119 / 0.299 \\
BNN & $\sigma^2=0.25$ & 5  & 0.000 / 0.000 / 0.180 & 0.000 / 0.004 / 0.253 & 0.018 / 0.130 / 0.317 \\
BNN & $\sigma^2=0.25$ & 10 & 0.000 / 0.000 / 0.156 & 0.000 / 0.012 / 0.231 & 0.019 / 0.135 / 0.305 \\
BNN & $\sigma^2=0.10$ & 3  & 0.000 / 0.000 / 0.131 & 0.000 / 0.008 / 0.211 & 0.019 / 0.116 / 0.301 \\
BNN & $\sigma^2=0.10$ & 5  & 0.000 / 0.000 / 0.141 & 0.000 / 0.006 / 0.210 & 0.020 / 0.125 / 0.299 \\
BNN & $\sigma^2=0.10$ & 10 & 0.000 / 0.000 / 0.155 & 0.000 / 0.012 / 0.199 & 0.020 / 0.143 / 0.284 \\
\hline
FNN & dropout = 0.20 & 3  & 0.007 / 0.020 / 0.205 & 0.004 / 0.017 / 0.195 & 0.031 / 0.111 / 0.275 \\
FNN & dropout = 0.20 & 5  & 0.021 / 0.038 / 0.206 & 0.014 / 0.039 / 0.197 & 0.038 / 0.110 / 0.280 \\
FNN & dropout = 0.20 & 10 & 0.020 / 0.020 / 0.198 & 0.014 / 0.021 / 0.197 & 0.041 / 0.111 / 0.275 \\
FNN & dropout = 0.50 & 3  & 0.010 / 0.015 / 0.184 & 0.005 / 0.021 / 0.189 & 0.031 / 0.110 / 0.273 \\
FNN & dropout = 0.50 & 5  & 0.026 / 0.072 / 0.204 & 0.021 / 0.067 / 0.197 & 0.047 / 0.110 / 0.276 \\
FNN & dropout = 0.50 & 10 & 0.047 / 0.082 / 0.246 & 0.038 / 0.073 / 0.258 & 0.056 / 0.136 / 0.279 \\
\hline
\end{tabular}
\end{table*}


\subsection{Results from smaller training and testing data}

We considered the same neural networks for a much smaller dataset including 8,000 training data and 2,000 testing data, drawn randomly from the original 70,000 and 20,000 training and testing data, respectively. The performance metrics are reported in Tables \ref{tab:fnn_summary_8k} and \ref{tab:bnn_summary_8k}. We see a drop in performance when we move to the smaller dataset, as expected, and worse performance as we move from $D_3$ to $D_2$ and then $D_1$. For frequentist training results, we generally see a drop in the AUPRC as we increase the number of layers $N_L$, with the exception of when using a dropout of 0.20 and $N_L=10$ for $D_2$. Brier scores remain similar across neural network structure, with larger scores for the less informative designs. A similar story occurs for the ECE values. These results are consistent for the 10 neuron case, but we do see an improved AUPRC for the 10 neuron models when $N_L=5$.

The Bayesian framework performs better on average than the frequentist when it comes to AUPRC score, with the exception of $D_2$, where the frequentist approach out performs for certain designs. Brier scores are generally lower for $D_3$ and $D_2$ in the Bayesian framework, with scores for $D_1$ comparable to frequentist. The Bayesian framework tends to do better across all metrics when 10 neurons are used in contrast to 20 neurons, though this improvement is relatively small.

\begin{figure*}
    \centering
    \includegraphics[width=0.8\linewidth]{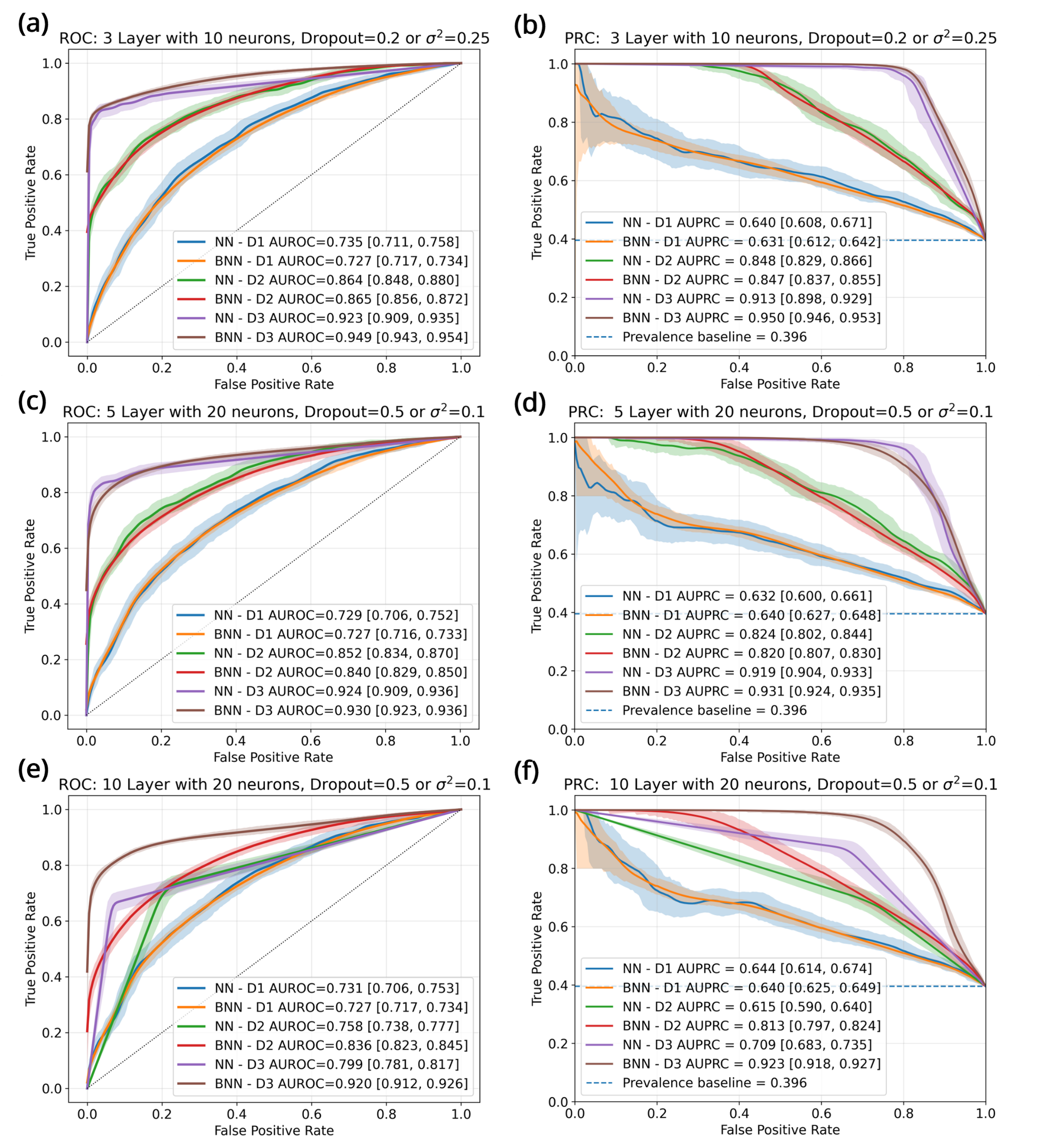}
    \caption{(Left) Receiver Operating Characteristic (ROC) curve for neural networks under the frequentist and Bayesian paradigm. Frequentist ROCs include 95\% confidence intervals while Bayesian ROCs include 95\% credible intervals, both also including the mean. Training data include 8,000 measurements, and results are shown for the 2,000 testing data points. (Right) Precision-Recall Curves (PRCs) for the corresponding ROC results.}
    \label{fig:ROC_PRC_8k}
\end{figure*}

The ROC and PRC curves are presented in Figure \ref{fig:ROC_PRC_8k}, which again illustrate a slight advantage towards the Bayesian framework. Note that the prevalence of SHD in this smaller dataset is 40\%, introducing a slight data imbalance. Is is evident that the uncertainty bounds are much larger in this data scenario. The frequestist approach is better than the Bayesian posterior mean in terms of AUC for $D_1$, while the Bayesian methods perform better for $D_2$ and $D_3$. We again see that shallower networks (3-Layer) tend to out perform deeper networks across the board. We also observe quite large uncertainty for the 10-layer network, likely due to the increased flexibility of the model. For the Bayesian ROC curves, we see larger uncertainty in the possible ROCs in comparison to the results in Figure \ref{fig:ROC_PRC_70k}, highlighting the role of sample size in the uncertainty of these methods.

Then risk-coverage results are provided in Table \ref{tab:combined_20n_risk_only_8k_6_27_67} for the smaller dataset. We again see that Bayesian results are better than the frequestist for smaller coverage values. The improvement is slightly more drastic for the smaller dataset at 6\% and 27\% coverage, with results for 67\% coverage again showing that frequentist networks can achieve lower risk under certain settings.

\begin{table*}[htbp]
\centering
\caption{Summary performance metrics for frequentist neural networks trained with 8K samples and evaluated on 2K test samples. Columns are grouped by neuron count and specified as (dropout, number of layers ($N_L$)). Best values for each metric across all configurations are shown in bold. Higher AUPRC is better, while lower Brier score and ECE are better.}
\label{tab:fnn_summary_8k}
\begin{tabular}{lcccccc}
\hline
Metric | (Dropout,$N_L$)
& (0.20, 3) & (0.20, 5) & (0.20, 10)
& (0.50, 3) & (0.50, 5) & (0.50, 10) \\
\hline
\multicolumn{7}{c}{\textbf{20 neurons}} \\
\hline
AUPRC D3
& 0.9265 & 0.8988 & 0.8798
& 0.9226 & 0.8887 & 0.8613 \\
AUPRC D2
& 0.8528 & 0.7944 & 0.8227
& 0.8484 & 0.7642 & 0.7661 \\
AUPRC D1
& 0.6284 & 0.6013 & 0.6214
& 0.6227 & 0.6037 & 0.6279 \\
\hline
Brier D3
& 0.1098 & 0.1134 & 0.1084
& 0.1108 & 0.1094 & 0.1129 \\
Brier D2
& 0.1602 & 0.1638 & 0.1627
& 0.1596 & 0.1655 & 0.1669 \\
Brier D1
& 0.2162 & 0.2198 & 0.2150
& 0.2171 & 0.2172 & 0.2116 \\
\hline
ECE D3
& 0.1923 & 0.1824 & 0.1914
& 0.1890 & 0.1890 & 0.1834 \\
ECE D2
& 0.1258 & 0.1148 & 0.1269
& 0.1246 & 0.0939 & 0.1134 \\
ECE D1
& 0.1007 & 0.1115 & 0.1031
& 0.1065 & 0.1149 & 0.0850 \\
\hline
\multicolumn{7}{c}{\textbf{10 neurons}} \\
\hline
AUPRC D3
& 0.9160 & 0.9100 & 0.7967
& 0.9293 & 0.9056 & 0.7789 \\
AUPRC D2
& 0.8484 & 0.8009 & 0.7460
& 0.8469 & 0.8029 & 0.6480 \\
AUPRC D1
& 0.6440 & 0.6049 & 0.6241
& 0.6276 & 0.6284 & 0.5891 \\
\hline
Brier D3
& 0.1110 & 0.1109 & 0.1258
& 0.1078 & 0.1103 & 0.1325 \\
Brier D2
& 0.1651 & 0.1651 & 0.1666
& 0.1655 & 0.1631 & 0.1832 \\
Brier D1
& 0.2151 & 0.2179 & 0.2194
& 0.2168 & 0.2188 & 0.2268 \\
\hline
ECE D3
& 0.1883 & 0.1874 & 0.1539
& 0.1938 & 0.1879 & 0.1391 \\
ECE D2
& 0.1281 & 0.1278 & 0.1039
& 0.1141 & 0.1231 & 0.0762 \\
ECE D1
& 0.1042 & 0.1079 & 0.1162
& 0.1049 & 0.1091 & 0.1394 \\
\hline
\end{tabular}
\end{table*}

\begin{table*}[htbp]
\centering
\caption{Summary performance metrics for Bayesian neural networks trained with 8K samples and evaluated on 2K test samples. Columns are grouped by neuron count and specified as (prior variance, number of layers ($N_L$)). Best values for each metric across all configurations are shown in bold. Higher AUPRC is better, while lower Brier score and ECE are better.}
\label{tab:bnn_summary_8k}
\begin{tabular}{lcccccc}
\hline
Metric | ($\sigma^2$, $N_L$)
& (0.25, 3) & (0.25, 5) & (0.25, 10)
& (0.10, 3) & (0.10, 5) & (0.10, 10) \\
\hline
\multicolumn{7}{c}{\textbf{20 neurons}} \\
\hline
AUPRC D3
& 0.9506 & 0.9495 & 0.9387
& 0.9307 & 0.9351 & 0.9274 \\
AUPRC D2
& 0.8503 & 0.8433 & 0.8359
& 0.8204 & 0.8267 & 0.8215 \\
AUPRC D1
& 0.6434 & 0.6416 & 0.6383
& 0.6472 & 0.6469 & 0.6475 \\
\hline
Brier D3
& 0.0689 & 0.0694 & 0.0767
& 0.0879 & 0.0829 & 0.0860 \\
Brier D2
& 0.1435 & 0.1433 & 0.1532
& 0.1557 & 0.1536 & 0.1561 \\
Brier D1
& 0.2164 & 0.2166 & 0.2196
& 0.2160 & 0.2155 & 0.2146 \\
\hline
ECE D3
& 0.0302 & 0.0346 & 0.0408
& 0.0333 & 0.0320 & 0.0312 \\
ECE D2
& 0.0789 & 0.0655 & 0.1020
& 0.0748 & 0.0761 & 0.0756 \\
ECE D1
& 0.1227 & 0.1131 & 0.1129
& 0.1190 & 0.1181 & 0.1077 \\
\hline
\multicolumn{7}{c}{\textbf{10 neurons}} \\
\hline
AUPRC D3
& 0.9528 & 0.9532 & 0.9437
& 0.9286 & 0.9377 & 0.9277 \\
AUPRC D2
& 0.8529 & 0.8474 & 0.8418
& 0.8210 & 0.8269 & 0.8240 \\
AUPRC D1
& 0.6397 & 0.6453 & 0.6419
& 0.6471 & 0.6463 & 0.6461 \\
\hline
Brier D3
& 0.0668 & 0.0641 & 0.0714
& 0.0890 & 0.0798 & 0.0893 \\
Brier D2
& 0.1426 & 0.1463 & 0.1467
& 0.1554 & 0.1538 & 0.1534 \\
Brier D1
& 0.2146 & 0.2181 & 0.2192
& 0.2164 & 0.2194 & 0.2144 \\
\hline
ECE D3
& 0.0359 & 0.0208 & 0.0400
& 0.0348 & 0.0231 & 0.0557 \\
ECE D2
& 0.0795 & 0.0868 & 0.0812
& 0.0746 & 0.0779 & 0.0636 \\
ECE D1
& 0.1132 & 0.1216 & 0.1119
& 0.1200 & 0.1341 & 0.1013 \\
\hline
\end{tabular}
\end{table*}

    
\begin{table*}[htbp]
\centering
\caption{Selective-risk comparison for 20-neuron Bayesian (BNN) and frequentist (FNN) neural networks trained with 8K samples and evaluated on 2K test samples. Entries are reported in the order D3 / D2 / D1. Lower risk indicates fewer errors among the retained least-uncertain predictions at the stated coverage. Bayesian models are indexed by prior variance $\sigma^2$, and frequentist models by dropout rate. Coverage levels are approximately 6\%, 27\%, and 67\%.}
\label{tab:combined_20n_risk_only_8k_6_27_67}
\small
\begin{tabular}{llcccc}
\hline
Framework & Tuning & Layers & Risk @ 6\% & Risk @ 27\% & Risk @ 67\% \\
\hline
BNN & $\sigma^2=0.25$ & 3  & 0.000 / 0.000 / 0.236 & 0.002 / 0.011 / 0.237 & 0.029 / 0.128 / 0.296 \\
BNN & $\sigma^2=0.25$ & 5  & 0.000 / 0.000 / 0.211 & 0.004 / 0.017 / 0.226 & 0.034 / 0.144 / 0.266 \\
BNN & $\sigma^2=0.25$ & 10 & 0.000 / 0.000 / 0.138 & 0.015 / 0.034 / 0.198 & 0.045 / 0.171 / 0.255 \\
BNN & $\sigma^2=0.10$ & 3  & 0.000 / 0.000 / 0.203 & 0.011 / 0.040 / 0.298 & 0.047 / 0.144 / 0.306 \\
BNN & $\sigma^2=0.10$ & 5  & 0.000 / 0.000 / 0.211 & 0.013 / 0.036 / 0.281 & 0.047 / 0.139 / 0.301 \\
BNN & $\sigma^2=0.10$ & 10 & 0.000 / 0.000 / 0.163 & 0.013 / 0.045 / 0.220 & 0.053 / 0.134 / 0.264 \\
\hline
FNN & dropout = 0.20 & 3  & 0.008 / 0.000 / 0.122 & 0.011 / 0.026 / 0.222 & 0.045 / 0.107 / 0.284 \\
FNN & dropout = 0.20 & 5  & 0.016 / 0.057 / 0.301 & 0.032 / 0.109 / 0.258 & 0.057 / 0.125 / 0.295 \\
FNN & dropout = 0.20 & 10 & 0.008 / 0.024 / 0.220 & 0.024 / 0.058 / 0.224 & 0.058 / 0.097 / 0.279 \\
FNN & dropout = 0.50 & 3  & 0.008 / 0.024 / 0.187 & 0.013 / 0.030 / 0.234 & 0.044 / 0.108 / 0.284 \\
FNN & dropout = 0.50 & 5  & 0.008 / 0.098 / 0.350 & 0.013 / 0.151 / 0.305 & 0.054 / 0.130 / 0.279 \\
FNN & dropout = 0.50 & 10 & 0.033 / 0.122 / 0.195 & 0.049 / 0.128 / 0.226 & 0.071 / 0.132 / 0.273 \\
\hline
\end{tabular}
\end{table*}

\subsection{Output uncertainty and triage: towards clinically relevant outputs}

\begin{figure*}
    \centering
    \includegraphics[width=0.8\linewidth]{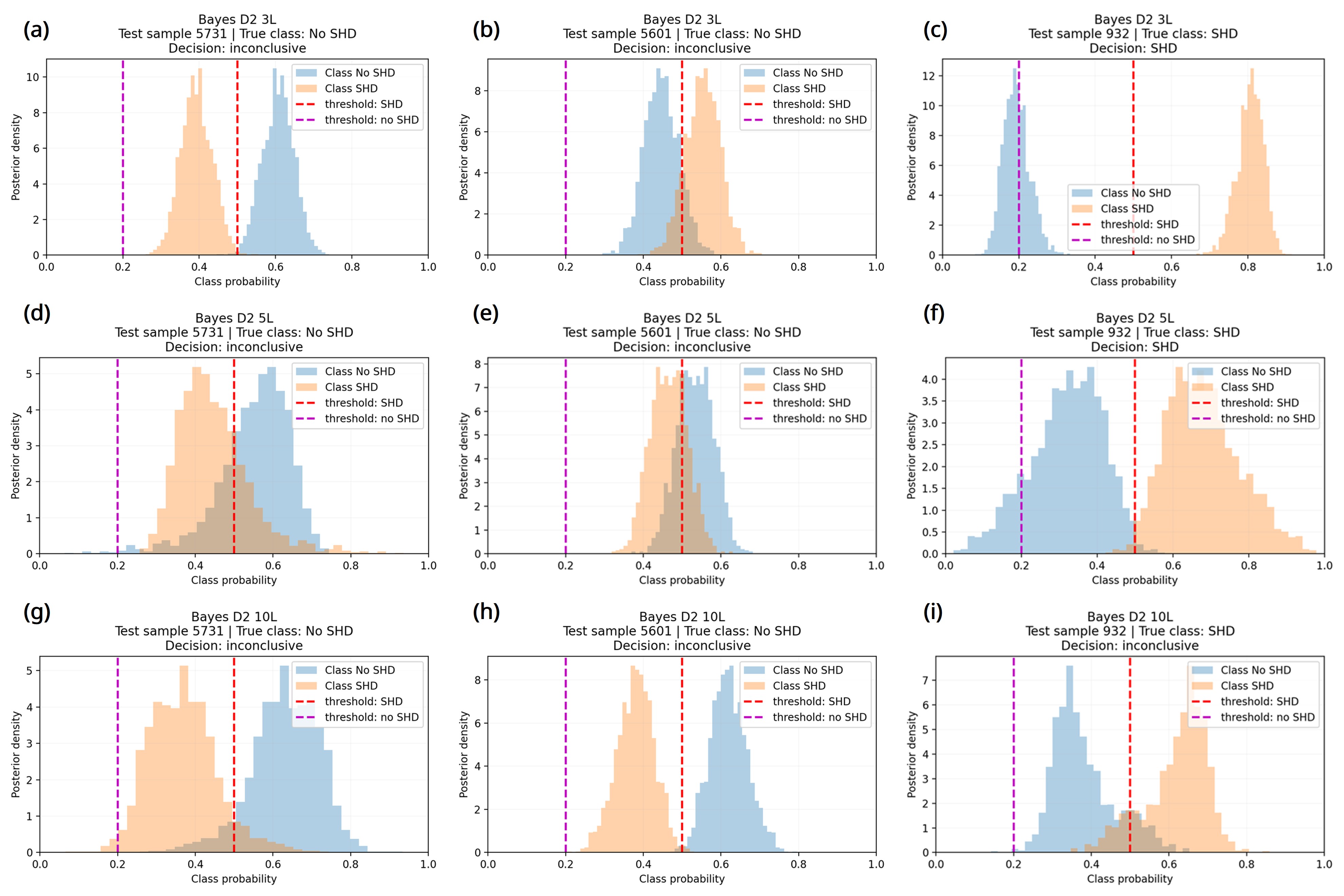}
    \caption{Posterior predictions over the probabilities for test data from the 70,000 training and 20,000 testing dataset. Results are shown for $D2$, with specific thresholds for class decisions or ``inconclusive'' decisions (see main text). All models are neural network with 10 neurons, and a prior variance of $\sigma^2=0.25$. Results compare three layers (top row), five layers (middle row), and ten layers (bottom row) across three datasets with the highest variability between models. The cutoffs for inconclusive decisions are provided as dotted lines.}
    \label{fig:OuputUQ_70K}
\end{figure*}

Given that the design $D_3$ is an upper bound for performance and that $D_2$ does substantially better in prediction than $D_1$, we investigated how this design, which only includes two echocardiography metrics, can be used for a better understanding of predicting SHD and for building a triage system. Recall that our definition of ``inconclusive'' findings was given by the SHD prediction 95\% credible intervals including an upper bound beyond 0.2 or SHD lower bounds including 0.5. Figure \ref{fig:OuputUQ_70K} shows the posterior probabilities for four randomly selected test datasets using this criteria for dataset $D_2$, 10 neurons per layer, and a prior variance of $\sigma^2=0.10$ for the 3-, 5-, and 10-layer networks. These examples show how the shallower networks tend to have tighter predictive distributions, and that increasing network depth tends to widen the 95\% credible interval. Specifically looking at  Figure \ref{fig:OuputUQ_70K}(c), (f), and (i), we see how a correct prediction of SHD in 3- and 5-layer networks becomes ``inconclusive'' as the predictive distribution has higher variance, leading to a 95\% credible interval that bleeds over the prescribe threshold (0.2). For a different test dataset in Figure \ref{fig:OuputUQ_70K}(b), (e), and (h), we see that the 3-layer network has a posterior that favors the SHD class, but is deemed inconclusive because its of the overlapping posteriors. As we move towards the 5- and 10-Layer model, the true class (No SHD) becomes more probable, though we still predict inconclusive.

\begin{figure*}
    \centering
    \includegraphics[width=0.8\linewidth]{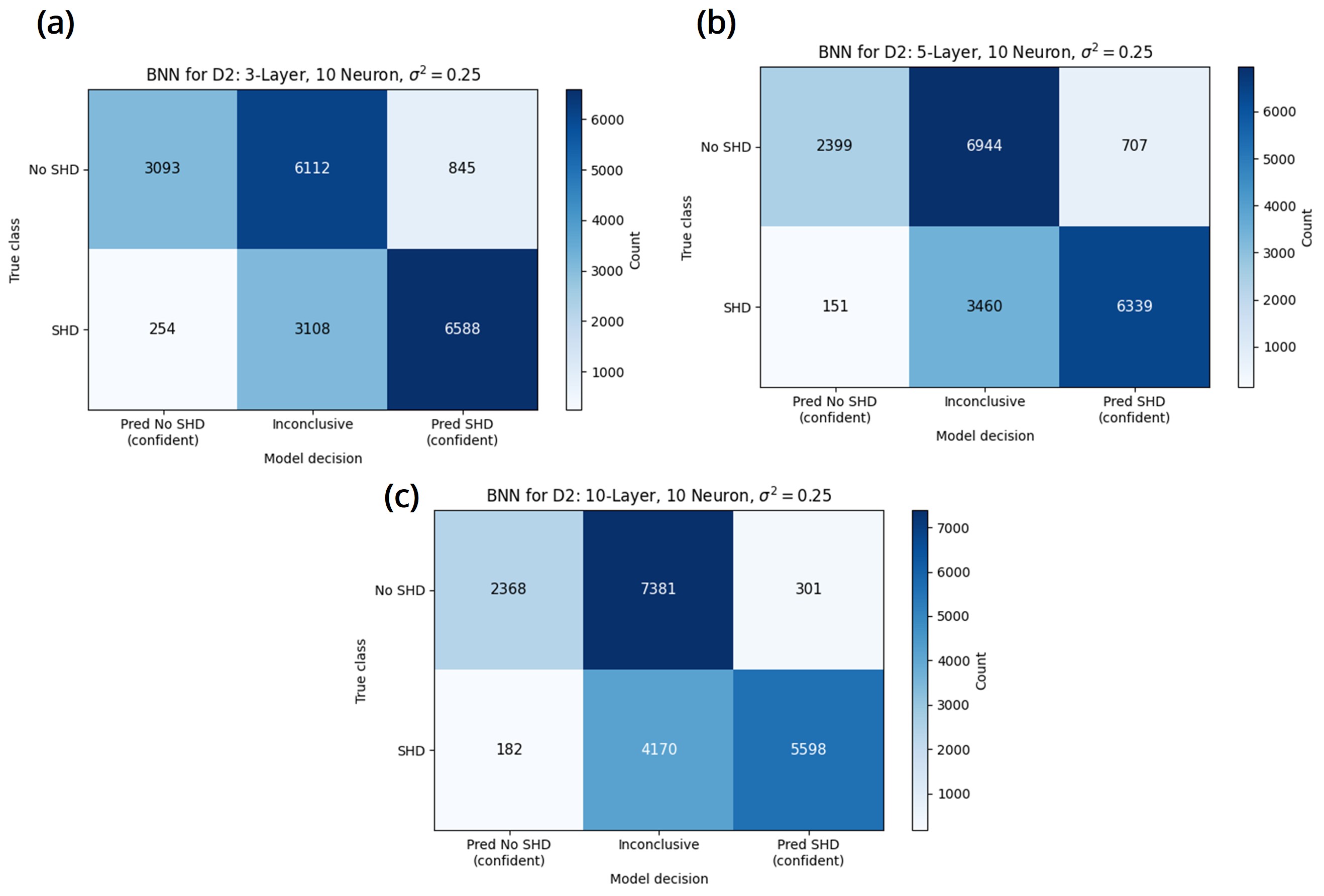}
    \caption{Confusion matrices for the (a) 3-layer, (b) 5-layer, and (c) 10-layer neural networks trained in a Bayesian framework with 10 neurons in each layer and a prior variance of $\sigma^2=0.25$. Predictions that do not satisfy the quantile rules described earlier are deemed ``inconclusive.''}
    \label{fig:Triage_70K}
\end{figure*}

Finally, we provide a test example of how this traige system would work if implemented through the confusion matrix presented in Figure \ref{fig:Triage_70K}. As mentioned, a cutoff of 0.2 for deeming an individual as not having SHD is extremely conservative, but reduces the chance of having a false negative. Our results show exactly this: the false positive rate is roughly 1\%. As a tradeoff, our framework is inconclusive for a majority of the samples. From a practical setting, this decision support system would prioritize inconclusive subjects over confident predictions, and reduce the immediate burden by roughly 50\% if expert sonographers were needed.

    
\section{Discussion}
We contrast frequentist and Bayesian approaches to training neural networks for the classification of SHD, providing some evidence in the power of the Bayesian methodology. While seemingly more mathematically complex, the Bayesian approach has clear benefits in its interpretation and handling of data, specifically in terms of uncertainty. While there are substantial amounts of uncertainty based methods within frequentist statistics, their application in the age of big-data approachs are more sparse and less utilized \cite{Kompa2021,Ding2020}. The interpretable Bayesian approach using probability distributions in the weight-space of the models provides a clearer interpretation of uncertainty in class predictions of the model. This can be especially useful in the case of noisy data, such as echocardiography.

Neural networks have long shown promise in the diagnosis and prediction of cardiovascular disease. Early work by Das et. al \cite{Das2009} used neural network ensembles to predict healthy and heart disease classes using blood samples and more ``generalized'' metrics from echocardiography measurements. The more recent study by Samad et al. \cite{Samad2019} compared machine learning models for survival prediction using increasingly complex datasets with (i) routine clinical and demographic data (e.g., sex, age, cholesterol levels, smoking status), (ii) added left ventricular EF estimates, and (iii) added full echocardiography reports. Samad found that age, TR Max Velocity, and left ventricular EF were some of the most important features, and that the average AUROC was 0.89 for 5-year mortality prediction. Prior studies have used Bayesian forms of machine learning for cardiovascular risk assessment as well. The study by Ordovas et al. \cite{Ordovas2023} used ``Bayesian Network Models,'' which are at their root a graphical model that encodes relationships between covariates for decision making, to predict the probability of hypertension and diabetes based on socioeconomic status, among other inputs.

Perhaps the most closely related study to our approach is the original analyses the led to ValveNet \cite{Elias2022} and EchoNet \cite{Poterucha2025}. The development of Valvenet by Elias et. al \cite{Elias2022} focused on the use of time-series electrocardiograms for the detection of left-sided valvular heart disease. Their model, which combined convolutional and residual neural network architectures, was able to classify and differentiate aortic stenosis, aortic regurgitation, and mitral regurgitation with AUROC values at or above 0.77 using ECG data alone. The follow up study by Porterucha et al. \cite{Poterucha2025} developed EchoNet and led to the publication of the EchoNext dataset (used in the present study). The EchoNet model was a similar architecture to ValveNet, but was transformed into a multiclass classifier for predicting multiple targets of SHD, including low left ventricular EF, tricuspid regurgitation, right ventricular systolic dysfunction, and elevated TR Max Velocity. The authors found overwhelming evidence that ECG data alone has strong predictive power in detecting specific forms of SHD.

Our approach is more simplistic in which data are included from the EchoNext dataset, but also provide new insight. First, while $D_1$ is the worst performing of the three datasets, it only contains the summary metrics from the full ECG signal as used previously \cite{Elias2022,Poterucha2025}. While a roughly 0.7 to 0.73 AUROC is only fair in predictive power, it does show that the summary information from ECG is informative in screening some SHD subjects. We also see that including three routine echocardiography metrics, the presence of pericardial effusion, IVS, and PWT, provided a substantial increase in the predictive power of the model. This shows that measurements are easily calculated with minimal uncertainty can provide substantial increases in predictive power for SHD. The significant increases in AUROCs and AUPRCs for $D_2$ thus provide a reasonable design for sonographers who typically rely on expert input: if you can get a routine ECG, and provide a measure of pericardial effusion, IVS, and PWT (with or without additional machine learning assistance), then SHD can be screened at a relatively high rate of accuracy. While the data in $D_3$ is used as an upper bound for our comparison study, it is clear from the high accuracy of this dataset that EF and TR Max Velocity, when combined with the other metrics in $D_1$ and $D_2$, is extremely accurate in detecting SHD. This is no surprise, as both of these metrics are used in the screening of SHD by expert sonographers \cite{Elias2022,Poterucha2025}. 

Bayesian approaches are not new by any means \cite{Friedman1997,Arbel2026,Elsayad2015}, though they are less commonly used in the era of deep-learning models. Part of this reason is because Bayesian approaches have typically requires Markov chain Monte Carlo (MCMC) methods to draw samples from the posterior. These methods do not scale well computationally when the parameter dimensionality is large \cite{Arbel2026}, hence their use with deep neural networks is limited, even when using cutting edge methods like Hamiltonian Monte Carlo \cite{Jospin2022}. The relatively recent implementation of variational inference methods in freely available source code (e.g., Pyro and NumPyro \cite{bingham2019pyro,phan2019composable}, used here) has reduced this computational burden. Our results using variational inference provide a slight, but consistent, edge over frequentist based inference. The most clear evidence of this is the Brier and ECE scores and the risk-coverage metrics as shown in Tables \ref{tab:fnn_summary_70k}, \ref{tab:bnn_summary_70k}, \ref{tab:fnn_summary_8k}, and \ref{tab:bnn_summary_8k}, and in Tables \ref{tab:combined_20n_risk_only} and \ref{tab:combined_20n_risk_only_8k_6_27_67}, respectively. For the large training set of 70,000 samples with 20,000 test data, the Brier scores are smaller for the Bayesian approach in $D_3$ and $D_2$, reflecting better calibrated predictive means for the classes, while these metrics are similar between frequentist and Bayesian methods for $D_1$. While this is a notable improvement, the Bayesian ECE scores outperform frequentist by an order of magnitude across all designs and most neural network parameterizations. This is somewhat expected, since the average of the posterior predictive naturally incorporates uncertainty in a more robust manner than in the frequentist networks.  The risk assessment in Table \ref{tab:combined_20n_risk_only} shows that the Bayesian framework consistently provides smaller risk values across the higher confidence data (6\% and 27\%), regardless of network architecture, for $D_3$ and $D_2$. As mentioned, the data in $D_1$ provides little to no information about heart function outside of pacing, and thus provides minimal data, owing to its larger risk, ECE, and Brier scores. This sort of analysis is extremely pertinent in healthcare applications, given that healthcare oriented machine learning models with zero tolerance for errors will require prediction-uncertainty metrics and identifying appropriate thresholds \cite{Chua2022}. A full comparison of AUROC, AUPRC, ECE, Brier, and risk-coverage estimates are necessary in comparison classification models. The work by Ding et al. \cite{Ding2020} emphasizes this point, showing that models that predict incorrectly more often can still obtain comparable AUROC and AUPRC, while their risk-coverage curves can provide better information about model trustworthiness.

We compare these models under the case of smaller training and testing data to reflect how these methodologies extend when data is either sparse for a specific disease case or where data cannot be shared, and thus require a ``internal'' dataset that does not have as many samples. We see many of the same trends as were the case in the large training set, but the magnitudes of the metrics have changed (for the worse). We still observe a slight preference towards Bayesian methods in terms of AUPRC, though this advantage is small. The Brier and ECE scores are substantially smaller in $D_3$ for the Bayesian approach, but this gap shrinks for $D_2$. Frequentist trained networks actually perform better in terms of Brier and ECE metrics for $D_1$. When it comes to risk-coverage assessments, Bayesian methods consistently outperform for $D_3$ across all presented coverages. For $D_2$, which is the design we are most interested in clinically, we see mixed results: Bayesian methods provide lower risk, on average, up to the 27\% coverage, but frequentist methods provide lower risk at higher coverage. This suggests that Bayesian methods might be preferred when data are considered ``trust worthy'' only. Nevertheless, we find sufficient evidence here to suggest that, even for small datasets such as those collected locally within a hospital system, a Bayesian approach is appropriate when developing machine learning classification tools.

Finally, we leveraged the use of the Bayesian framework and the predictive distributions to consider the case of triage systems for echocardiography screening. This application is especially well suited for machine learning, as broad classes of AI algorithms have made an impact on helping non-experts (such as nurses or healthcare providers with little ultrasonography experience) acquire echocardiography information from patients \cite{Fazlalizadeh2024}. While this could be done with frequentist training, the Bayesian approach provides a more interpretable result for the posterior predictive distribution for a given class, as illustrated in Figure \ref{fig:OuputUQ_70K}. The choice of threshold for decision making is non-trivial, and should be attributed to the disease type, practicing medical personnel, and data noise. The studies highlighted in \cite{Fazlalizadeh2024} showcase how clinicians can work synergistically with machine learning protocol development. Our results provide one of these approaches, where we only want to make decisions on subject data when we are highly confident, based on the 95\% predictive posterior, that individuals fall into a specific class. In the setting of echocardiography triage, as shown in Figure \ref{fig:Triage_70K}, setting a lower bound of 20\% for ruling out SHD leads to a relatively robust screening that minimizes confident false negatives. As a consequence, a majority of data is deemed inconclusive, which requires expert insight. If this pipeline was used in practice, roughly 7,000 - 9,000 of the test data (35\% to 45\%) would be confidently and correctly classified and removed from the triage list, while approximately 9,000 to 11,000 of test subjects (45\% to 55\%) would be prioritized for expert sonography. Even a 30-40\% reduction in the number of individuals require expert sonographer analysis could have massive impacts on cardiologists availability to serve their patient population.

There are several limitations of the proposed work. We compared frequentist and Bayesian approaches across similar neural network architectures, including width, depth, dropout, and prior variance. Though width and depth were comparable, we opted to use ReLU activation functions in frequentist settings and hyperbolic tangents when we use Bayesian inference. This was to help simulate the most common approaches use in the literature, since ReLU mitigates the vanishing gradient problem in standard machine learning, but can cause issues in the Bayesian setting \cite{Castillo2025}. This study focuses on tradeoffs and possible improvements in using the Bayesian approach, but additional hyperparameter tuning (e.g., learning rate, unequal network width, activation function choice) could be conducted to improve predictive performance. We used three designs in this study, with $D_3$ including multiple echocardiographic metrics and serving as a ``upper bound'' for training performance. These metrics were already identified through the EchoNext project \cite{Elias2022,Poterucha2025}, hence future work should focus on how Bayesian methods compare when starting from raw ECG and echocardiography signals. The training data used may include multiple recordings from the same individual, while the test data is unique across all subjects. This may lead to a slight bias during training. Finally, the authors of the EchoNext dataset provide full ECG waveforms as additional data \cite{Poterucha2025}. A next step in our analysis will include Bayesian inference paradigms applied to ECG data, and may provide even greater improvement in SHD classification as has been documented in other works \cite{Pedroso2026,Poterucha2025,Ulloa-Cerna2022}.

\section{Conclusion}
We present a comparative analysis of frequentist and Bayesian neural network classifiers on a dataset combining ECG and echocardiography data. Our findings show that Bayesian neural networks showed comparable or modestly improved discrimination and substantially improved calibration and selective-risk behavior under the tested architectures. We show that this trend holds true even in the case of significantly smaller amounts of training data, owing to the Bayesian framework's ability to appropriately quantify uncertainty. We show that this architecture can be useful in the form of echocardiography triage, and provide a proof-of-concept for future work aimed at optimizing healthcare resources, especially in the setting of rural healthcare.

\section*{CRediT authorship contribution statement}
\textbf{Mitchel J. Colebank}: Conceptualization, Methodology, Software, Formal Analysis, Investigation, Resources, Visualization, Funding Acquisition, Writing - Original Draft, Writing - Review \& Editing.

\section*{Declaration of competing interest}
The author declares that he has no known competing financial interests or personal relationships that could influence the work reported in this paper.

\section*{Acknowledgements}
This work was funded in part by a South Carolina EPSCoR Grant for Applications in Industry and Networking Program (26-GA01), through the University of South Carolina's NIAID-funded R25AI164581 (R25) Big Data Health Science Fellow Program, and through an American Heart Association Career Development Award (26CDA1599801).

\section*{Data sharing statement}
The EchoNext dataset is available through PhysioNet \cite{PhysioNet,goldberger2000physiobank} at \url{https://physionet.org/content/echonext/1.1.0/}. The code that trains the neural networks and runs the plotting scripts can be found at \url{https://github.com/mjcolebank/Colebank_ECG_Echo_Triage}.




    
\bibliographystyle{cas-model2-names}
\bibliography{refs}

\end{document}